\documentclass[aps,pre,superscriptaddress,nofootinbib,preprintnumbers]{revtex4-2}

\usepackage{graphicx}
\usepackage{graphics}
\usepackage{amsmath,amssymb}
\usepackage[usenames]{color}
\usepackage{subfigure}
\usepackage{hyperref}
\usepackage{enumitem}
\usepackage{lgreek}
\usepackage{float}

\newcommand{\be}{\begin{equation}}
\newcommand{\ee}{\end{equation}}
\newcommand{\bea}{\begin{eqnarray}}
\newcommand{\eea}{\end{eqnarray}}
\newcommand{\ben}{\begin{enumerate}}
\newcommand{\een}{\end{enumerate}}
\newcommand{\bit}{\begin{itemize}}
\newcommand{\eit}{\end{itemize}}

\newcommand{\la}[1]{\label{#1}}

\newcommand{\vv}[1]{\mathbf #1}							

\newcommand{\bert}{\raise-0.45mm\hbox{\Large$\Box$}}			

\makeatletter
\newcommand*\bigcdot{\mathpalette\bigcdot@{.5}}
\newcommand*\bigcdot@[2]{\mathbin{\vcenter{\hbox{\scalebox{#2}{$\m@th#1\bullet$}}}}}
\makeatother

\definecolor{BrickRed}{cmyk}{0,0.89,0.94,0.28}					
\definecolor{MidnightBlue}{cmyk}{0.98,0.13,0,0.43}				
\definecolor{DarkGreen}{rgb}{0.100806,0.495968,0.209979}
\definecolor{orange}{rgb}{0.587167,0.354498,0.146197}

\graphicspath {{figures/}}

\begin{document}

\title{Thermal Vacuum Cosmology Explains Hubble Tension }
\author{Robert Alicki}
\email{robert.alicki@ug.edu.pl}
\affiliation{International Centre for Theory of Quantum Technologies (ICTQT), University of Gda\'nsk,  80-308, Gda\'nsk, Poland}

\date{\today}

\begin{abstract}
It is argued that the previously proposed modification of the standard (flat) inflationary $\Lambda CDM$ model  in which cosmological constant  is replaced by  thermal energy of  expanding vacum, characterized by the  Gibbons-Hawking temperature, explains the origin of notorious ``Hubble tension''.

\end{abstract}

\maketitle

The Hubble constant ($H_0$) is a very important cosmological parameter, directly related to the current expansion rate of our Universe, its composition, and ultimate fate. However, the $H_0$ values
deduced from observations of the early and the late Universe do not agree with each other creating the so-called “Hubble tension”  which has become a hot topic in modern
cosmology of the last decades (see e.g. \cite{Perlmutter} - \cite{Sunny} and references therein). In particular, one can compare two well-established cosmological probes : 1) type Ia supernovae (SN Ia) data based on direct measurements for the local Universe , 2) cosmic microwave background (CMB) radiation data, used for higher redshifts and  assuming \emph{the validity of the $\Lambda$CDM standard model}.
\par
Generally, the data for local  Universe favor a higher value $H_0 \simeq 74$ while the CMB-based data give a lower value
$H_0 \simeq 68$ ( in standard units [$km s^{-1} Mpc^{-1}$]). A tension of almost 10\% is essentially larger than the expected accuracy of the recent  \emph{precision cosmology} data which reaches 1\%
\cite{Turner}.
\par
The aim of this Letter is to replace  the standard $\Lambda$CDM model by the recently developed Thermal Vacuum Model (TVM) \cite{QTdS} -\cite{TVM} \footnote{Notice the change of notation and units to a more common in the cosmological literature.} in order to explain the Hubble tension. The TVM is based on the standard Friedmann equations for the flat Friedmann-Lema\^itre-Robertson-Walker (FLRW) metric $ds^2 = -c^2 \ dt^2 + a^2 (t) \, d \vv x^2$  written in terms of the Hubble parameter  $H = \dot{a}/a$ and in SI units
\be
 \frac{3}{8\pi G} H^2 = \frac{1}{c^2} \rho , 
\la{eq:Fried1}
\ee
\be
 \dot{H} = - \frac{3}{2} H^2  + \frac{4\pi G}{c^2} p , 
\la{eq:Fried2}
\ee
where $\rho$ and $p$ are energy density and pressure of the ``cosmic fluid''. The main assumption  of TVM (see \cite{QTdS},\cite{relax},\cite{TVM}) is the decomposition of the energy density
\be
\rho = \rho_m + \rho_{dS}
\la{eq:energy}
\ee
 where the first term is the energy density of matter (including Standard Model and Dark Matter particles) and the second is the  thermal vacuum ``dark energy''. 
 \par
 Similarly,  pressure  
\be 
 p = w_m \rho_m - \rho_{dS}
 \la{eq:pressure}
\ee
depends on the standard state equation for matter $w_m$ and reflects the property of  dark energy which is not diluted by space expansion  ($w_{dS}= -1$). 
\par
The TV energy density $\rho_{dS}$ depends on the Hubble parameter $H(t)$ and varies in time in contrast to  the cosmological constant $\Lambda$ of the standard  $\Lambda$CDM model. It is interpreted as energy density of the expanding vacuum which is treated as thermal equilibrium state of all degrees of freedom describing visible and dark matter. As argued in \cite{QTdS},\cite{relax} this equilibrium state is completely characterized by the  temporal Gibbons-Hawking temperature  \cite{Gibbons}
\be 
T_{dS}(t)= \hbar H(t) /2\pi k_B . 
 \la{eq:GHtemp}
\ee 
This theory applied to the very early Universe provides a new mechanism of inflation and its graceful exit accompanied by particle production (at the expence of TV energy density), without introducing  inflaton field and reheating mechanism. The TVM applied to the late Universe  describes  observed acceleration of expansion. Moreover, this formalism has been combined with the anomalous quantum gravity effects leading to  a viable baryogenesis mechanism  and certain bounds on possible dark matter models \cite{TVM}. 
\par
According to the TVM the expanding vacuum of the very late Universe can be modelled by a cold gas of massive elementary particles at  the particular density corresponding to a single particle,  of a given polarization, occupying the  volume  $\lambda_{\rm th}^3( T_{dS})$ defined by  the thermal de Broglie wavelength. This ``dark energy'' density is given by (see \cite{TVM} eqs.(18) - (21))
\be
 \rho_{dS}(t) = \frac{3 c^2}{8\pi G} [{H_{\infty}}]^{1/2}  [{H}(t)]^{3/2} , 
\la{eq:edark}
\ee
where $H_{\infty}$  is an ``ultimate Hubble constant'' predicted by the TVM and dependent on the mass spectrum for elementary particles of ordinary and dark matter.
Inserting \eqref{eq:edark} into the TVM Friedmann equations for the flat Friedmann-Lema\^itre-Robertson-Walker Universe 
one obtains the following formula for the redshift-dependent Hubble parameter (see \cite{TVM} eq.(26))
\be
 H_{TV}(z) = H_0 \left[\left(1 -\Omega_{TV}\right) (1 +z)^{3/4} + \Omega_{TV} \right]^2 ,
\la{eq:hTVM}
\ee
where $1>\Omega_{TV} > 0$ is given by  $\Omega_{TV}= \sqrt{H_{\infty}/H_0}$.
\par
On the other hand, for the  $\Lambda$CDM model applied to the late Universe the analogical formula reads
\be
 H_{\Lambda} (z) =  H_0 \sqrt{(1- \Omega_{\Lambda}) (1 + z)^3 + \Omega_{\Lambda}} ,
\la{eq:hLCMD}
\ee
where $ \Omega_{\Lambda}$ is the energy density of cosmological constant relative to the critical density. 
\par
To compare these two expresions it is convenient to use a new dimensionless function
\be
 f(z) \equiv  \frac{H(z)}{H_0 (1 + z)},
\la{eq:fz}
\ee
with the property that its local minimum defines the redshift $z_{ac}$ at which accelerated expansion begins. Fig.1 shows the function $f(z)$ for both models, with a generaly accepted value $ \Omega_{\Lambda} =0.7$ and with  $\Omega_{TV} =  0.42 $ choosen to reproduce the same redshift $z_{ac}\simeq 0.7$.

\begin{figure}[t]
	\includegraphics[width=0.7 \textwidth]{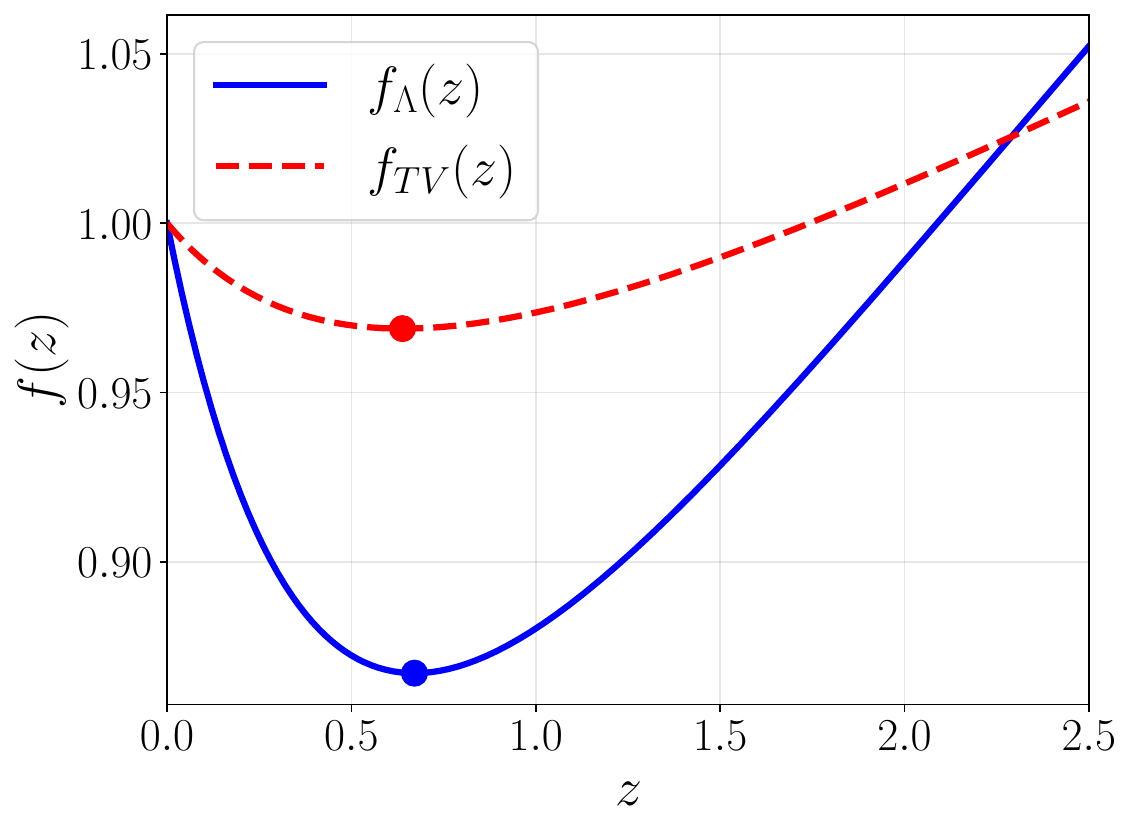}
\caption{Plot of the $f(z)$ curves:  for  $\Lambda$CDM model with $ \Omega_{\Lambda} =0.7$,  TVM with $\Omega_{TV} = 0.42 $. The dots mark minima at $z_{ac} \simeq 0.7$. }
\end{figure}

One can test both formulas  for $H(z)$, \eqref{eq:hTVM}, \eqref{eq:hLCMD}, using the following observables :

a)  luminosity distance  $D_L(z)$ as a function of  redshift  (``standard candles'' method),
\be
 D_L(z) = c(1+z) \int_0^z \frac{dz'}{H(z')},
\la{eq:dL}
\ee

b)  angular diameter distance $D_A (z) $ determined by the angle $\Theta$ subtended by an object of known physical length $R$ (``standard ruler``)
\be
 D_A = \frac{R}{\Theta},\quad  D_A (z) = \frac{ D_L(z) }{(1+z)^2}.
\la{eq:dA}
\ee
Firstly, one can consider the case of local Universe  ($z << 1$) and derive, using \eqref{eq:dL}, \eqref{eq:hTVM} and \eqref{eq:hLCMD}, the second order expression for  $D_L(z)$ which has an identical form for both models
\be
 D_L(z)\simeq \frac{c z}{H_0} \left[ 1 + \frac{1}{4}(1 + 3\Omega) z\right],
\la{eq:dLapp}
\ee 
with  $\Omega = \Omega_{\Lambda}$ or $\Omega_{TV}$. It is clear ( see e.g. \cite{Hubble1} Fig.1, for observational data) that for the local Universe the difference between both models appearing in the quadratic correction to the linear Hubble law is too small to be detected within the present day accuracy. The linear fit for small $z$ gives $H_0 \simeq 74$.
\par
To explain the source of the  Hubble tension one  assumes that the formula \eqref{eq:hTVM} based on the TVM  is correct and can be compared with the measured observables given by \eqref{eq:dL} or \eqref{eq:dA}. Then the formulas based on  the $\Lambda$CDM expression for  $H(z)$ \eqref{eq:hLCMD} cannot be correct and must be modified by replacing $H_0$ with  the ``running Hubble constant'' $\hat{H}_0(z)$ defined by the following identity
\be
\frac{D_L (z)}{c(1+z)}=\frac{1}{H_0}\int_0^z \frac{d z'}{\left[\left(1 -\Omega_{TV}\right) (1 +z')^{3/4} + \Omega_{TV} \right]^2}
\equiv \frac{1}{\hat{H}_0(z)}\int_0^z \frac{d z'}{\sqrt{(1- \Omega_{\Lambda}) (1 + z')^3 + \Omega_{\Lambda}}},
\la{eq:RunH}
\ee
(see \cite{RunHubble} and references therein for a different but related idea of ``effective running Hubble constant'').
The values of $\hat{H}_0(z)$ are obtained by computing numerically the ratio of two integrals and inserting the  value of $H_0$.

\par
\begin{figure}[t]
	\includegraphics[width=0.7 \textwidth]{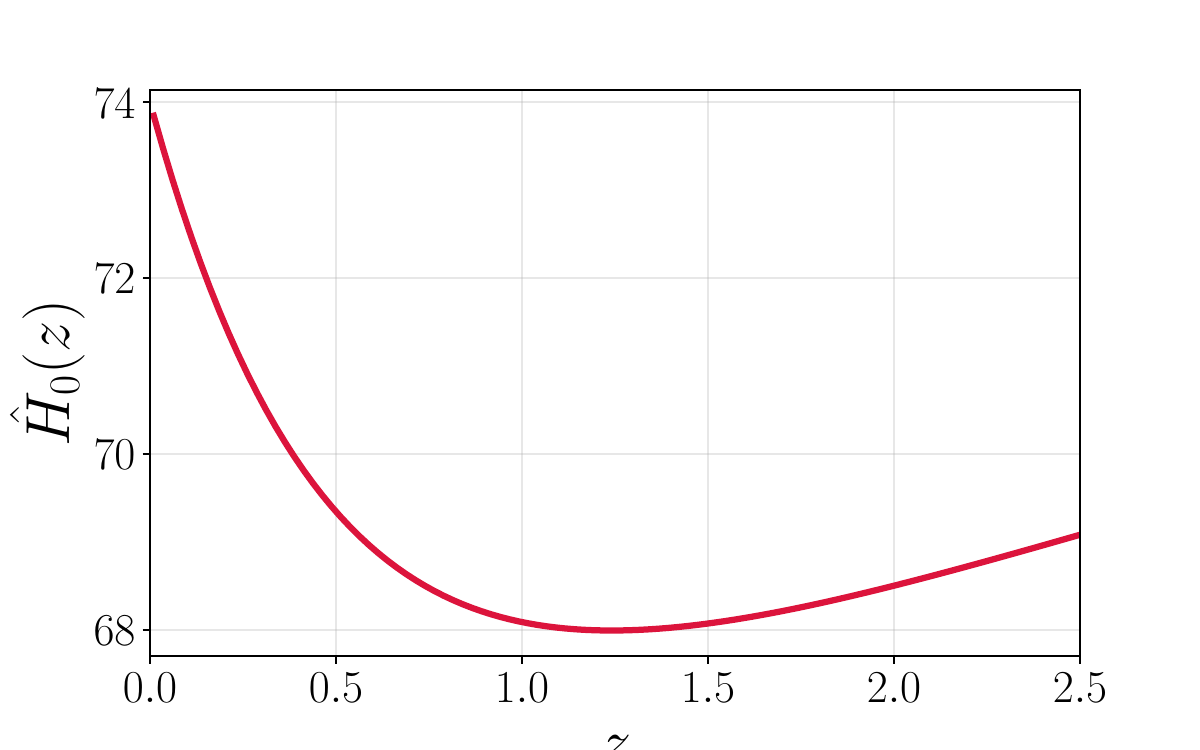}
\caption{Plot of the running Hubble constant  $\hat{H}_0(z)$ for $\Omega_{TV} = 0.42 ,  \Omega_{\Lambda} =0.7$ , and  $H_0 = 74$ .}
\end{figure}

Fig.2. shows the plot of $\hat{H}_0(z)$, for $z\in[0, 2.5]$, with $ \Omega_{\Lambda} =0.7$ ,  $\Omega_{TV} =  0.42 $ and  $H_0 = 74$. The origin of  the  Hubble tension is clearly visible. For $z << 1$ the running Hubble constant is well approximated by the local Universe value $H_0 \simeq 74$ while for $ 0.5 < z < 2.5$, the lower value $H_0 \simeq 68$ is recovered. It means that the application of the $\Lambda$CDM model to analyze the measurement results yields running Hubble constant, averaged over a particular range of  the redshift, while using TVM should produce the true $H_0\simeq 74$.
 This observation seems to be a strong argument for the validity of cosmology based on the Thermal Vacuum Model.

{\bf Acknowledgements}:  The author thanks Borhan Ahmadi for making figures .


\end{document}